\documentclass[preprint,prd,aps]{revtex4-1}
\usepackage{amsmath}
\usepackage{graphicx}
\usepackage{longtable}

\usepackage{color}

\begin{document}

\bibliographystyle{apsrev}

\title{Probing the gravitational Faraday rotation using quasar X-ray microlensing}


\author{Bin Chen}
\email{bchen3@fsu.edu}
\affiliation{Research Computing Center, Dirac Science Library, Room 150-F,
Florida State University, Tallahassee, FL 32306, USA}


\begin{abstract}
The effect of gravitational Faraday rotation 
was predicted in the 1950s, but there is currently no practical method for measuring this effect. 
Measuring this effect is important because it will provide new evidence for correctness of general relativity, in particular, in the strong field limit.
We predict that the observed degree and angle of the X-ray polarization of a cosmologically distant quasar microlensed by the random star field in a foreground galaxy or cluster lens vary rapidly and concurrently with flux during caustic-crossing events using the first simulation of quasar X-ray microlensing polarization light curves. 
Therefore, it is possible to detect gravitational Faraday rotation by monitoring the X-ray polarization of gravitationally microlensed quasars.
Detecting this effect will also confirm the strong gravity nature of quasar X-ray emission.
\end{abstract}

\maketitle

\section{Introduction}

Gravitational Faraday rotation, the rotation of the plane of polarization of an electromagnetic wave propagating in a curved spacetime, 
is the gravitational analogue of the electromagnetic Faraday rotation \cite{Balazs58, Plebanski60, Godfrey70, Connors77, Stark77, Piran85, Ishihara88, Agol97, Sereno05, Schnittman09}.
While this effect exists in both the strong and weak field regimes \cite{Darwin59,Virbhadra00, Bozza01,Sereno03,Bozza06}, significant gravitational Faraday rotations are expected only for strong gravitational field, for example, for photons transiting regions very close to a Kerr black hole (a black hole with spin \cite{Kerr63}).
Measuring this effect is difficult for at least two reasons: first, the regions producing significant rotations (a few gravitational radii from a back hole), are too small to be directly resolved by current telescopes; and the lack of polarimeters with high sensitivity which can measure polarization fraction at a few percent level and resolve polarization angle at a few degrees level.  
It is, however, important to detect this effect because it tests general relativity, in particular, in the strong field limit.
Recent gravitational microlensing observations have constrained the quasar X-ray emission sizes to be about 10 gravitational radii of the central supermassive black hole \cite{Kochanek04,Chartas09,Dai10,Chen11,Morgan12,Ana13,Blackburne15,Macleod15}.  
Gravitational Faraday rotation should leave its fingerprint on such compact regions around black holes.
In this work we show that it is possible to detect the effect of gravitational Faraday rotation by monitoring the X-ray polarization of gravitationally microlensed quasars.


For distances within a few $r_g$ (gravitational radius) of a Kerr blackhole, light paths are strongly bent by the gravity field of the black hole, and a photon's polarization plane will be rotated by the gravitational Faraday effect as it transits this region.
Consequently, the X-ray flux and polarization of a microlensed quasar at cosmological redshift $z_s$ will be lensed first by the supermassive black hole powering the active galactic nuclei (AGN), then microlensed by the foreground random star field in the lens galaxy at redshift $z_d$. 
Given the large cosmological distance from the lens galaxy to the background quasar, the strong lensing by the Kerr black hole can be decoupled from the foreground microlensing using the numerical schemes outlined in \cite{Chen13b}.   
The gravitational lensing by a foreground galaxy can produce multiple images with angular separations of the order of $1$ arc second.
A `macro' image of the background quasar generated by foreground galaxy lensing can be further microlensed by random stars in the lens galaxy close to the image position.
The image splitting by a stellar-mass gravitational lens is only at the micro arc second level, not resolvable by telescopes; however, the observed flux of a  microlensed quasar varies with time due to the relative motion between the lens, source, and observer, i.e., the microlensing light curve \cite{Chang79}.
As we will show, the X-ray polarization of a microlensed quasar also fluctuates with time, producing a microlensing polarization light curve strongly correlated with the classical flux light curve.


\section{Results}

Figure~\ref{fig:shear} shows a simulated microlensing magnification pattern by a random star field with mean lens mass $\langle M_*\rangle=0.3 M_\odot,$ 
external shear $\gamma=(0.2,0)$, surface mass density $\kappa_{\rm c}=0.4$ (smooth mass) and $\kappa_*=0.2$ (compact objects) \cite{Wambsganss99}.
For simplicity, we have assumed a log-uniform lens mass distribution with $0.01M_\odot<M<1.6M_\odot$.
The image size is about $5\theta_E\times 5 \theta_E$ where $\theta_E$ is the Einstein ring angle of the mean lens mass $\langle M_*\rangle.$ 
We have assumed a standard $\Lambda$CDM cosmology with $\Omega_{\rm m}= 0.3,$ $\Omega_{\Lambda}=0.7$ and Hubble constant $H_0= 70\, \rm km\, s^{-1} Mpc^{-1}.$
The diamond-like curves are microlensing caustics where the magnification diverges for point sources.
The inset shows the intensity profile (with size amplified by a factor of 8) of a compact X-ray source of radius $20\,r_g$ lensed by a supermassive black hole with mass $M_{\rm BH}=10^9M_\odot,$ spin $a=0.998,$ and inclination angle $\theta=75^\circ.$
We have assumed a simple corona geometry, a thin disk above the optical accretion disk moving with Keplerian flow \cite{Haardt91}.
The inset also shows the X-ray polarization profile as lensed by the Kerr black hole and twisted by the gravitational Faraday rotation generated by the Kerr black hole ray-tracing code `KERTAP' \citep{Chen15} (refer to the Methods Section for details). 
The unlensed polarization profile was given as in the classical work of Chandrasekhar \cite{Chandra60} 
for a scattering-dominated atmosphere. 
In the rest (comoving) frame of the source, the unlensed polarization is horizontally aligned (i.e., has a polarization angle $\chi=0$) and the polarization degree $\delta$ increases with the inclination angle ($\delta=0$ for the face on case and $\delta\approx 12\%$ for the edge on case).
The degree of polarization varies across the X-ray disk because of Einstein's light bending and aberration effects, and the polarization angle varies because of the gravitational Faraday rotation.
The lensed X-ray flux is strongly focused in a small region (in dark red) to the left of the black hole where the X-ray emitting source is approaching the observer, and the polarization is very small in this region. 
The total (averaged) polarization, about $1.6\%$, is much smaller than the unlensed case ($\delta =4.6\%$ for inclination angle $\theta=75^\circ$). 
The inset of Fig.~\ref{fig:shear} shows that gravitational Faraday rotation is only significant in regions very close to the central black hole.
Consequently, directly observing this gravitational twisting across an AGN X-ray corona is not possible due to its small size not resolvable by any contemporary X-ray telescope.

The idea of detecting gravitational Faraday rotation via quasar microlensing is very simple.
Given the small size of the quasar X-ray emission region,
 the observed X-ray polarization is the average/integral across the whole X-ray corona.
Since the polarization varies across the black-hole-lensed X-ray corona, the observed net X-ray polarization of a microlensed quasar is a convolution of the Kerr-lensed profile with the microlensing magnification pattern produced by the foreground random star field. 
Consequently, the microlensed quasar X-ray polarization should vary with time when the source moves across the microlensing magnification pattern due to relative transverse motions of the lens, source, and  observer (see, e.g., the dashed curves in Fig~\ref{fig:shear}) \cite{Agol96,Chen13b}. 
For example, the observed polarization fraction should decrease when the dark red region in the inset of Fig~\ref{fig:shear} is crossing a caustic because this region has a low polarization and its flux is greatly amplified by the microlensing, and consequently has a much larger weight when averaged across the disk.
Similarly the averaged polarization angle $\chi$ drops when this region is crossing a caustic given that the polarization is gravitationally rotated downward (clockwise, $\chi<0$) in this region.
The result is the opposite when the region on the right hand side of the black hole crosses a caustic, because this region has a larger than average polarization and an upwardly (counterclockwise) rotated polarization angle.
Therefore, we expect to observe rapid fluctuations in the X-ray polarization degree and angle when a compact X-ray source crosses a caustic. 
This variation should be strongly correlated with fluctuations in the observed X-ray flux.

Figures~\ref{fig:lc1} and \ref{fig:lc2} show the first simulation of quasar X-ray microlensing polarization light curves. 
The quasar moves along the dashed orange and red lines in Fig.~\ref{fig:shear} for Figs.~\ref{fig:lc1} and \ref{fig:lc2}, respectively.
For both examples we have chosen an X-ray disk of radius $20\,r_g$.
But for Fig.~\ref{fig:lc1} the black hole has spin $a=0.998$ and is observed at an inclination angle $\theta=75^\circ,$ whereas for Fig.~\ref{fig:lc2}
we have assumed spin $a=0.6$ and inclination $\theta=45^\circ.$
The inner cutoff of the X-ray disk is selected to be the innermost stable circular orbit in the equatorial plane ($r_{\rm ISCO}=1.24\,r_g$ and $3.83\,r_g$ for $a=0.998$ and $0.6,$ respectively \cite{Bardeen72}). 
The doubly (Kerr and micro) lensed  X-ray flux magnitude, and the polarization degree and angle, fluctuate rapidly and concurrently during caustic crossing.
For the first example (high inclination angle and large spin) the polarization angle $\chi$ can change up to $\pm3^\circ$ due to gravitational microlensing (the average gravitational-Faraday-rotated polarization angle is about $-8^\circ$ without microlensing). 
For the second example (smaller inclination angle and moderate black hole spin) the polarization angle can change up to $-6^\circ$ (the average polarization angle is about $-6^\circ$ without microlensing).
But for the second example the net polarization, $\delta\approx0.7\%,$ is less than one half of the first case $\delta\approx1.6\%$ (disk observed nearly edge on).
Consequently, one may detect gravitational Faraday rotation by a multiple-epoch observation of a microlensed quasar (such as Q~0957+561 and Q~2237+0305) in the X-ray band. 
Such an experiment requires a X-ray polarimeter with a minimum detectable polarization (MDP) at $\sim$$1\%,$ 
and a resolution of the polarization angle at $\sim$$1^\circ.$ 
The traditional polarimeters based on the Bragg diffraction technique \cite{Weisskopf78} do not have this sensitivity; however, with the new generation of X-ray polarimeters built upon the micro-pattern gas detector technique, a MDP at $\sim1\%$ is indeed possible provided that the integration time is long enough \cite{Costa01}.
Furthermore, these new X-ray polarimeters perform best in the soft X-ray band ($E<10\,\rm keV$), the energy band in which most of the X-ray photons from lensed quasars are observed. 
Therefore, our strategy for detecting the gravitational Faraday rotation is technically possible.

Monitoring the polarization of a microlensed quasar in the X-ray band for multiple epochs with long exposure times is probably expensive due to the lower photon count rates \cite{Chen12} compared with nearby objects such as MCG-6-30-15. 
As an alternative, since multiple images of a galaxy or cluster lens sit in different random star fields and the observed polarization of each image is the convolution of the source profile with an independent microlensing magnification pattern, one might observe different polarization degree and angle between images of a galaxy/cluster lens through an one-epoch observation (similar to the X-ray-to-optical flux ratio anomalies between image pairs caused by microlensing \cite{Pooley06}). 
Such a one-epoch observation requires a telescope being able to resolve multiple X-ray images of a lens (i.e., has an angular resolution similar to \emph{Chandra} X-ray Observatory, $\sim$0.5 arc second).
Since quasar optical/UV emission regions are larger than the X-ray emission regions by at least one order of magnitude \cite{Kochanek04,Chartas09,Dai10,Chen11,Morgan12,Ana13,Blackburne15,Macleod15}, the impact of gravitational Faraday rotation on the quasar optical/UV polarization should be less significant.
For example, the dilution of the observed net polarization should be less severe for optical emission. 
On the other hand, the foreground microlensing is also less significant because larger sources smooth out the effect of microlensing magnification more significantly.
Consequently, the fluctuations of the microlensing polarization light curve in the optical/UV band should be of smaller amplitude than those in the X-ray band.
This way of detecting the gravitational Faraday rotation by contrasting the X-ray and optical/UV polarization light curves requires a joint mission in both energy bands. 
Our method can also be applied to serendipitous discovery of the gravitational Faraday effect from Galactic/nearby X-ray sources microlensed by a random star (with high photon count rates but low chance of microlensing).

Black hole strong lensing and gravitational Faraday rotation significantly depolarize the AGN X-ray emission, making the already difficult polarization measurement even more challenging.
Fortunately, the foreground microlensing of a lensed quasar selectively magnifies different regions of the gravitational-Faraday-rotated X-ray emission, making it possible to detect this effect with the new generation of X-ray polarimeters \cite{Costa01} by converting the spacial variations/twisting of polarization into the time dependence of X-ray microlensing polarization light curves.
It has been more than 30 years since the last successful polarization measurements in the soft X-ray bands \cite{Weisskopf78}.
Recently, polarization detections were made in the Gamma-ray band for the Crab pulsar and Cygnus X-1 \cite{Dean08,Laurent11}.
It would be important to also have data from the X-ray polarimetry.
Measuring the polarization in the X-ray band should put strong constraints on emission models and AGN corona geometry. 
Since X-ray polarimetry missions are being continually proposed \cite{Costa11,Krawczynski12,Schnittman13}, our results should help in the process of designing these missions. 
If the predicted effect is observed, it will greatly enhance the scientific return of such campaigns by providing evidence of gravitational Faraday rotation and the strong gravity nature of AGN X-ray emission.  

\section{Discussion}

We have shown that the degree and angle of the X-ray polarization of a microlensed quasar vary rapidly and concurrently with flux during caustic-crossing events using a  simulation of quasar X-ray microlensing polarization light curves.
We have estimated the order of magnitude of fluctuations in the polarization fraction and angle caused by foreground microlensing using two simple examples based on a single microlensing simulation and a simple corona geometry (refer to \cite{Chen13a} for some other geometries).
The effect depends significantly on the parameters assumed, such as the black hole spin and disk inclination angle. 
A full statistical analysis requires an exploration of a large parameter space (for example, the corona geometry, emission profile, black hole mass and spin, and disk inclination angle for the background source; and mean mass, distribution of the random star field, external shear, and velocity of the relative motion for the foreground microlensing; see Tab.~1 of \cite{Chen13b}). 
Such a detailed analysis is beyond the scope of this paper but is highly desired. 

In our analysis we have used ray tracing to study the strong black hole lensing of the quasar X-ray emission \cite{Chen15}.
Analytical and semi-analytical formalism for black hole lensing exists and could have been used as an alternative to numerical ray tracing \cite{Virbhadra00,Bozza01}. 
We have assumed the standard weak field approximation for the foreground microlensing simulation, and have treated the microlensing stars as point masses without angular momenta. 
Since stars are spinning, their angular momenta contribute to the gravitational Faraday rotation of linearly polarized light passing by.
In particular, if the foreground microlenses are very compact (e.g., stellar-mass black holes), relativistic (ghost) images will be formed \cite{Darwin59} and the polarization of these relativistic images will be strongly modified by these lenses. 
Weak field approximation fails for such situations \cite{Bozza06}.  
Since relativistic images are very faint and the observed net polarization is an average weighted by image brightness, the contribution from the relativistic images produced by foreground microlensing stellar-mass black holes to the net gravitational Faraday rotation should be small.  
Furthermore, the gravito-magnetic effect of spinning stars perturbs the bending angle, image positions, caustics, and critical curves of the microlensing simulation \cite{Sereno03}. 
This will also introduce corrections to the gravitational Faraday rotation of the microlensed quasar X-ray emission.
It will be important to investigate them in future.




\section{Acknowledgements} 

The author thanks R. Kanowski and P. Beerli for reading of the manuscript. The simulation was generated on the HPC cluster at the Research Computer Center at the Florida State Univ.
 

\section{Author contributions statement} 

B.C. developed the code, generated the simulation, and wrote the paper.

\section{Additional information} \textbf{Competing financial interests} The author declares no competing financial interests.

\clearpage

\begin{figure*}
\begin{center}
\includegraphics[width=1.0\textwidth,height=0.5\textheight]{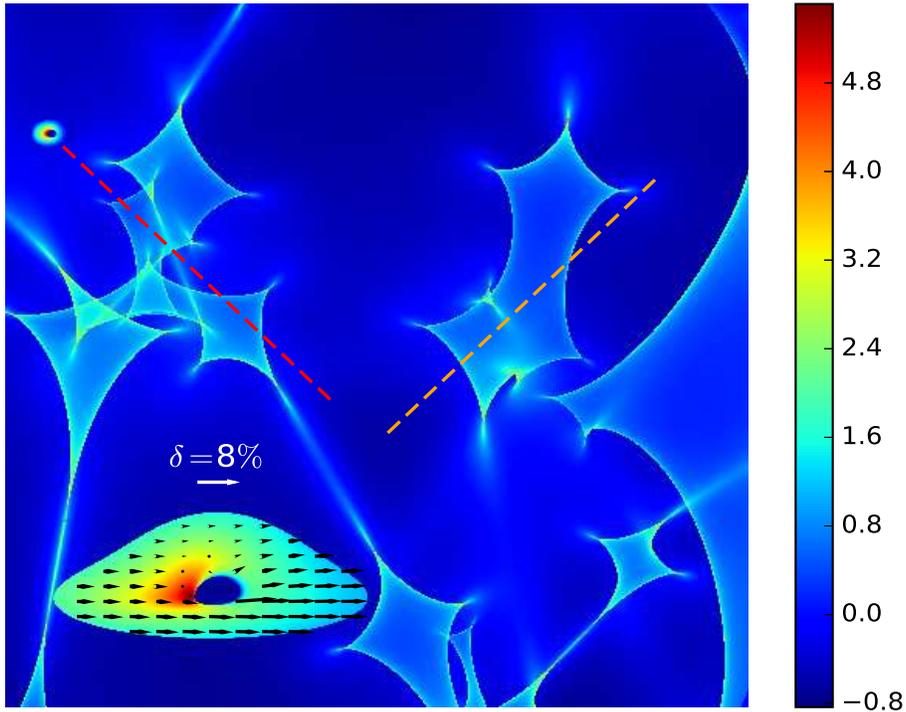}
\end{center}
\caption{Gravitational microlensing magnification pattern 
of a random star field with mean lens mass $\langle M_*\rangle=0.3 M_\odot$, external shear $\gamma=(0.2,0)$, surface mass density $\kappa_{c}=0.4,$ $\kappa_*=0.2$. 
The lens and source redshifts are $z_d=0.5$ and $z_s =1.0,$ respectively.
The image is of high resolution $8000 \times 8000$ and size about $5\theta_E\times 5\theta_E$ where $\theta_E$ is the Einstein ring angle of a lens of mass $\langle M_*\rangle$.
The inset is the Kerr strong lensing image of an X-ray disk of radius $20\,r_g$ by a black hole of spin $a = 0.998$ observed at an inclination angle $75^\circ.$ 
The arrows in the lensed image show the effect of black hole lensing and gravitational Faraday rotation on the X-ray polarization.
The color bar is in log scale.
}
\label{fig:shear}
\end{figure*}

\begin{figure*}
\begin{center}
\includegraphics[width=1.0\textwidth,height=0.5\textheight]{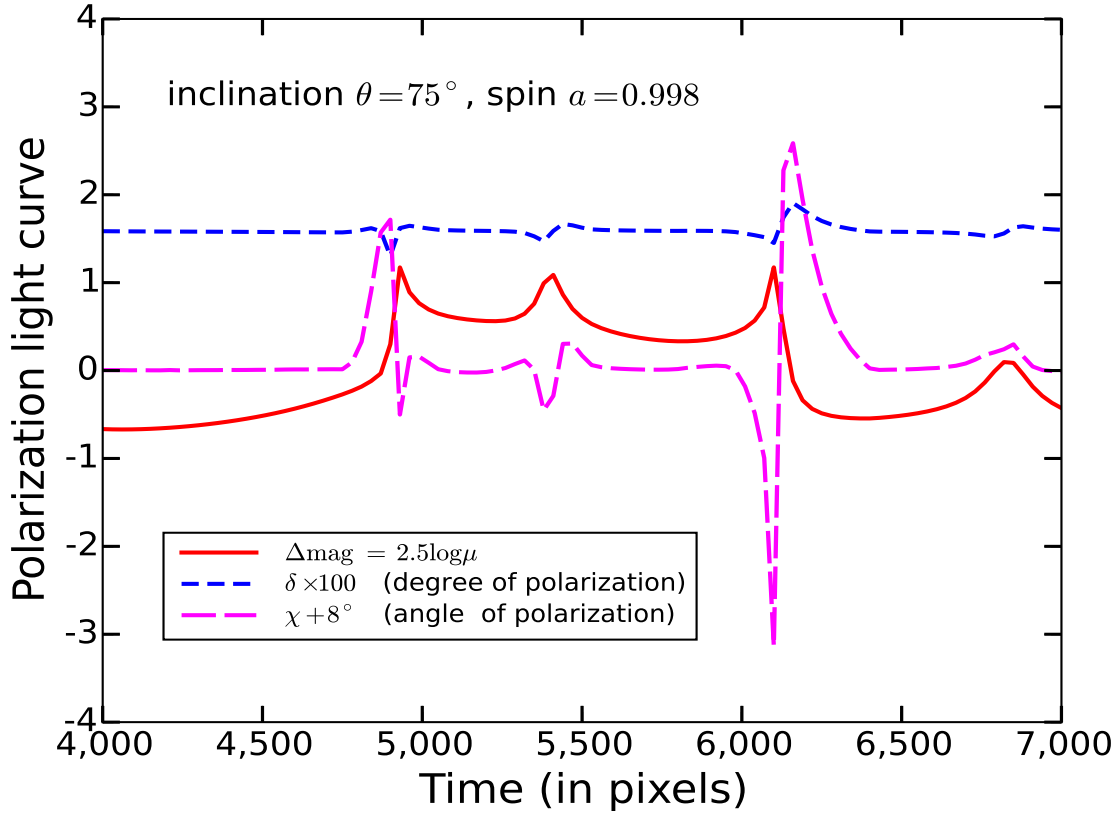}
\end{center}
\caption{Microlensing polarization light curve.
The black hole spin $a=0.998,$ and the accretion disk inclination angle $\theta=75^\circ.$
The trajectory of the quasar is shown as the dashed orange line in Fig.~\ref{fig:shear}.
The time is measured in pixels in the $x$ direction (one pixel is about $6.8$ days assuming the source is moving with a transversal velocity about $500\rm\, km\, s^{-1}$).
The red solid, magenta long-dashed, and blue dashed curves show respectively the variation of the X-ray flux magnitude (i.e., the traditional microlensing light curve), the angle $\chi,$ and the degree $\delta$ of the X-ray polarization. 
The observed X-ray flux, and the degree and angle of the X-ray polarization vary rapidly and concurrently during microlensing caustic crossing.
}
\label{fig:lc1}
\end{figure*}

\begin{figure*}
\begin{center}
\includegraphics[width=1.0\textwidth,height=0.5\textheight]{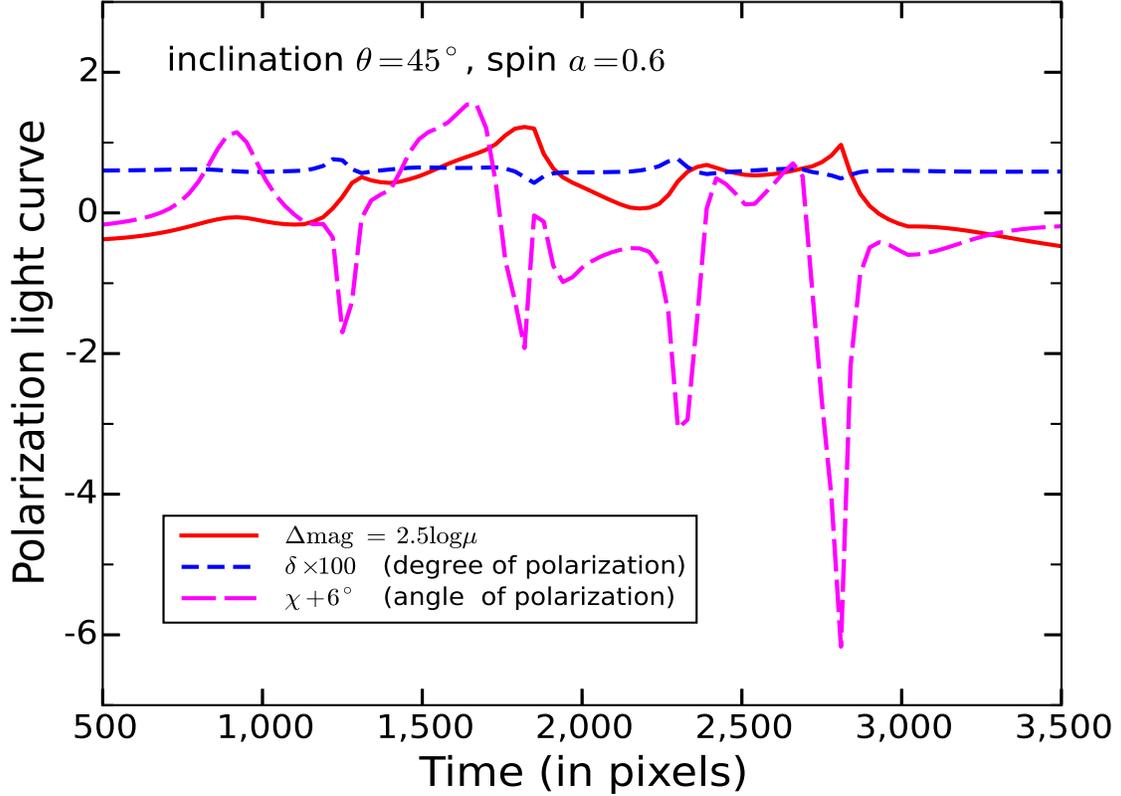}
\end{center}
\caption{ Microlensing polarization light curve.
The trajectory of the quasar is shown as the dashed red line in Fig.~\ref{fig:shear}.
The black hole spin $a=0.6,$ and the accretion disk inclination angle $\theta=45^\circ.$
The averaged polarization $\delta$ is only about $0.7\%$ (the unlensed $\delta$ increases with inclination, and $\delta=0$ for the face on case because of the symmetry).
The polarization angle can be gravitational Faraday rotated by up to $-6^\circ$ for this case.
Consequently, an X-ray polarimeter of higher resolution in polarization degree $\delta$ (better than $0.5\%$) is needed to detect the gravitational Faraday rotation of the X-ray polarization if the accretion disk is observed at a smaller inclination angle.   
}
\label{fig:lc2}
\end{figure*}

\clearpage

\section{Methods}

The algorithms for studying the effects of Kerr black hole strong lensing on quasar X-ray microlensing (not including polarization) were presented in \cite{Chen13b}.
A Kerr black hole ray-tracing code including gravitational Faraday rotation is available \cite{Chen15}.
The simulation in this work combined the algorithms developed by us and collaborators \cite{Chen13b, Chen15}.
We outline the essential steps in the following.

The microlensing simulation (Fig.~1) was generated using the formalism in \cite{Wambsganss99}. 
An important difference from traditional microlensing simulations is that standard microlensing simulations assume the source plane to be flat (no spacetime curvature); however, the ambient spacetime of the quasar X-ray corona is strongly curved because of the gravity field of the central supermassive black hole.  
Consequently, we have chosen the source plane to be a plane (orthogonal to the line of sight) between the background quasar and the foreground lens galaxy (at a distance $d\sim$$10^6\, r_g$ from the quasar black hole; $d\ll d_{ds}$ where $d_{ds}$ is the angular diameter distance from the lens to the source).
Instead of using the intrinsic (i.e., before lensed by the quasar black hole) source profile of the quasar X-ray corona,  we use the black-hole-lensed image of the corona as the source input for the foreground microlensing simulation.
In this way, we have decoupled the foreground microlensing ray-tracing from the background Kerr ray-tracing and greatly simplified the simulation scheme (refer to Fig.~1 of \cite{Chen13b}).  

To generate the intensity and polarization profile of an X-ray corona strongly lensed by the supermassive black hole including effects of gravitational Faraday rotation (i.e., the inset of Fig.~1), we have used the black hole ray-tracing code KERTAP \cite{Chen15}.
KERTAP is a backward black hole ray-tracing code solving the geodesic equations in Kerr spacetime using a $5^{\rm th}$ order Runge-Kutta algorithm with an adaptive step size control. 
This code can generate the image of an AGN X-ray corona lensed by a Kerr black hole with input parameters such as size of the X-ray corona, black hole spin, and the inclination angle of the accretion disk. 
It includes effects of the gravitational red- or blueshift, light bending, image distortion, multiple images, and gravitational Faraday rotation simultaneously.
Since the physics of AGN X-ray coronae is still poorly understood, we have chosen a simple disk geometry for the corona with its inner cutoff the innermost stable circular orbit \cite{Bardeen72} and an outer radius $20\, r_g$ based on constraints from recent quasar X-ray microlensing observations.
We have assumed a simple power-law spectrum (photon index $\Gamma=2$) for the intrinsic X-ray emission with steep radial profile $n=3$ (see \cite{Chen15} for details).
Since accretion disks are commonly believed to be optically thick, we have turned off the higher order images when generating the black-hole-lensed corona image.
After both the corona image and microlensing magnification pattern were generated, polarization light curves (Figs.~\ref{fig:lc1} and \ref{fig:lc2})  were generated by moving the Kerr-lensed corona image along a selected trajectory (e.g., dashed curves in Fig.~\ref{fig:shear}) in the magnification pattern and convolving the Kerr-lensed intensity and polarization profiles with the microlensing magnification pattern.

\end{document}